\documentclass[onecolumn,useAMS,usenatbib]{mnras}
\usepackage{graphicx}
\usepackage{subfigure}
\usepackage{xcolor}
\usepackage{mathtext,bbm,amsmath,amsfonts,amssymb,indentfirst,syntonly,graphicx}
\usepackage{mathtools}
\usepackage{slashbox}
\usepackage[english]{babel}
\usepackage{calc}
\usepackage{tikz}
\usepackage[T1]{fontenc}
\usepackage{ae,aecompl}
\usepackage{latexsym,graphicx,bbm}
\usepackage{enumerate}
\usepackage{longtable}
\usepackage{supertabular}

\title[Revised constraints on photon mass from FRBs]{Revised constraints on the photon mass from well-localized fast radio bursts}
\author[H.-N. Lin, L. Tang and R. Zou]
{Hai-Nan Lin$^{1,2}$, Li Tang$^{3}$\thanks{Corresponding author: tang@mtc.edu.cn} and Rui Zou$^{1,2}$\\
$^{1}$Department of Physics, Chongqing University, Chongqing 401331, China\\
$^{2}$Chongqing Key Laboratory for Strongly Coupled Physics, Chongqing University, Chongqing 401331, China\\
$^{3}$Department of Math and Physics, Mianyang Normal University, Mianyang 621000, China\\}
\begin{document}

\date{Accepted 2022; Received 2022; in original form 2022}

\pagerange{\pageref{firstpage}--\pageref{lastpage}} \pubyear{2022}

\maketitle

\label{firstpage}

\begin{abstract}
  We constrain the photon mass from well-localized fast radio bursts (FRBs) using Bayes inference method. The probability distributions of dispersion measures (DM) of host galaxy and intergalactic medium are properly taken into account. The photon mass is tightly constrained from 17 well-localized FRBs in the redshift range $0<z<0.66$. Assuming that there is no redshift evolution of host DM, the $1\sigma$ and $2\sigma$ upper limits of photon mass are constrained to be $m_\gamma<4.8\times 10^{-51}$ kg and $m_\gamma<7.1\times 10^{-51}$ kg, respectively. Monte Carlo simulations show that, even enlarging the FRB sample to 200 and extending the redshift range to $0<z<3$ couldn't significantly improve the constraining ability on photon mass. This is because of the large uncertainty on the DM of intergalactic medium.
\end{abstract}


\section{Introduction}\label{sec:introduction}

The invariance of the speed of light, which states that the speed of light in vacuum is a constant with respect to any inertial frame, is one of the two principles of Einstein's special relativity. A direct inference of this principle is that photon is massless. The non-zero photon mass, no matter how small it is, would imply new physics beyond the special relativity. Therefore, strictly constraining photon mass is of great importance. To this end, several methods, ranging from particle physics to cosmology scales, have been proposed to put tight constraint on photon mass. For instance, experimentally test of Coulomb's law \citep{Williams:1971ms}, investigate the gravitational deflection of electromagnetic radiation \citep{Lowenthal:1973ka}, measure the magnetic field of Jupiter \citep{Davis:1975mn}, test the Ampere's law by observing large-scale magnetic fields in galaxies \citep{Ryutov:2010zz}, observation of the spin of supermassive black hole \citep{Pani:2012vp}, measure the spin-down rate of pulsars \citep{Yang:2017ece}, and so on.

In the cosmological aspect, one possible way to constrain photon mass is to measure the frequency-dependent velocity of electromagnetic waves from astrophysical transients, see e.g. \citet{Wei:2021vvn} for recent review. If photon has non-zero mass, then electromagnetic waves of different frequencies travel with different velocities, hence we can observe relative time delay if non-monochromatic electromagnetic waves propagate from cosmological scale. For example, \citet{Lovell:1964} analyzed the time delay between optical and radio emissions from flare stars, and obtained an upper limit on the photon mass of $m_\gamma < 1.6\times 10^{-45}$ kg. \citet{Warner:1969} analyzed the optical emission of different wavelength from the Crab Nebula pulsar, and obtained $m_\gamma < 5.2\times 10^{-44}$ kg. \citet{Schaefer:1998zg} analyzed the time delay between gamma-ray emission and radio afterglow from gamma-ray bursts (GRBs), and obtained $m_\gamma < 4.2\times 10^{-47}$ kg. \citet{Zhang:2016ruz} investigated the radio afterglows and multi-band radio peaks from a large sample of GRBs, and obtained the strictest constraint $m_\gamma < 1.062\times 10^{-47}$ kg from GRB 050416A. \citet{Wei:2018pyh} investigated the dispersion measures of a sample of extragalactic radio pulsars, and put robust limit on photon mass with $m_\gamma < 1.51\times 10^{-48}$ kg.

Fast radio bursts (FRBs), as milliseconds radio transients happening in the Universe, are ideal astrophysical sources to constrain photon mass. The use of FRBs to constrain photon mass was first proposed by \citet{Wu:2016brq}, who obtained a stringent upper limits on photon mass $m_\gamma < 5.2\times 10^{-50}$ kg by analyzing the frequency-dependent time delays of FRB 150418. The result of \citet{Wu:2016brq} is based on the assumption that the time delay is purely caused by the non-zero photon mass. This is of course not reasonable, because a large part, if not all of the time delay is induced by the dispersion of photon by electrons. As is known, the dispersion-induced time delay and the time delay caused by non-zero photon mass have the same frequency dependence ($\Delta t\propto \nu^{-2}$), thus these two parts is indistinguishable using a single FRB. \citet{Shao:2017tuu} pointed out that, although the two parts of time delay have the same frequency dependence, they have remarkably different redshift dependence. Therefore, the degeneracy between them can be broken using a large sample of FRBs at different redshift. \citet{Shao:2017tuu} developed a Bayesian framework to constrain photon mass from a sample of FRBs with and without redshift measurement, and obtained $m_\gamma < 8.7\times 10^{-51}$ kg. Later on, \citet{Wei:2020wtf} used the same method to constrain photon mass with nine well-localized FRBs, and obtained $m_\gamma < 7.1\times 10^{-51}$ kg.

There are still some difficulties in using FRBs to constrain photon mass. The dispersion measure (DM) of an extragalactic FRB consists of several parts (see the next section for details), among which the DM of host galaxy is poorly known. Too many factors may affect the host DM, such as the galaxy type, the mass of host galaxy, the separation of FRB source from galactic center, the inclination angle of host galaxy, etc. Previous work often treated the host DM as a constant parameter \citep{Shao:2017tuu}, or parameterized it tracing the star formation rate \citep{Wei:2020wtf}. The constant assumption is of course inappropriate, as the host DM may vary significantly from bursts to bursts. On the other hand, \citet{Lin:2022afm} found no strong evidence for the correlation between host DM and star formation rate. A more proper way to deal with host DM is to consider the probability distribution and marginalize over it \citep{Macquart:2020lln,Zhang:2020mgq}. Another difficulty is that it is hard to calculate the DM of intergalactic medium (${\rm DM_{IGM}}$) due to the matter fluctuation of the Universe. Previous work often calculated ${\rm DM_{IGM}}$ using the mean matter density, and introduced an uncertainty term to account for the possible deviation from the mean \citep{Shao:2017tuu,Wei:2020wtf}. This treatment is based on the underlying assumption that the probability distribution of ${\rm DM_{IGM}}$ is Gaussian. However, theoretical analysis and numerical simulations shown that the actual value of ${\rm DM_{IGM}}$ may deviate from the mean significantly, and the probability distribution is non-Gaussian, which has a flat tail at large value \citep{Macquart:2020lln,Zhang:2020xoc}. The probability distribution of ${\rm DM_{IGM}}$ may affect the constraints on photon mass, which remains to be further investigated.

As more well-localized FRBs are observed recently, in this paper we will use them to constrain the photon mass, by properly taking into account the probability distribution of DMs of host galaxy and IGM. The structure of this paper is arranged as follows: The theoretical method is introduced in Section \ref{sec:methodology}. The observed FRBs and the constraining results on photon mass are presented in Section \ref{sec:results}. In Section \ref{sec:simulations}, we use Monte Carlo simulations to forecast the constraining ability on photon mass if more well-localized FRBs at high redshift are observed in the future. Finally, discussion and conclusions are given in Section \ref{sec:conclusions}.

\section{Methodology}\label{sec:methodology}

The speed of electromagnetic waves propagating in cold plasma is frequency-dependent, making low-frequency waves travel slower than high-frequency waves. This effect, although is small, may be detectable if it accumulates at cosmological scale. The relative time delay between low- and high-frequency electromagnetic waves propagating from a distant source to earth is given by \citep{Bentum:2016ekl,Wei:2018pyh}
\begin{equation}\label{eq:Delta_t_DM}
  \Delta t_{\rm DM}=\frac{e^2}{8\pi^2 m_e\epsilon_0 c}(\nu_l^{-2}-\nu_h^{-2}){\rm DM_{astro}},
\end{equation}
where $e$ and $m_e$ are the charge and mass of electron respectively, $c$ is the speed of light in vacuum, $\epsilon_0$ is the permittivity of vacuum, $\nu_l$ and $\nu_h$ are the low and high frequencies of electromagnetic waves respectively. ${\rm DM_{astro}}=\int n_edl$ is the dispersion measure (DM), which equals to the integral of electron number density along the line-of-sight.

On the other hand, the non-zero photon mass will also cause frequency-dependent speed of electromagnetic waves. The relative time delay between low- and high-frequency electromagnetic waves induced by the non-zero photon mass, in the condition of $m_\gamma c^2\ll h\nu$, can be written as \citep{Wu:2016brq,Shao:2017tuu}
\begin{equation}\label{eq:Delta_t_gamma}
  \Delta t_{m_\gamma}=\frac{1}{2H_0}\left(\frac{m_\gamma c^2}{h}\right)^2H_\gamma(z)(\nu_l^{-2}-\nu_h^{-2}),
\end{equation}
where $H_0$ is the Hubble constant, $h$ is the Planck constant, $m_\gamma$ is the photon mass, and $z$ is the cosmic redshift of the wave source. In the standard $\Lambda$CDM cosmological model, the quantity $H_\gamma(z)$ is given by
\begin{equation}\label{eq:H_gamma}
  H_\gamma(z)=\int_0^z\frac{1}{(1+z)^2}\frac{dz}{\sqrt{\Omega_M(1+z)^3+\Omega_\Lambda}},
\end{equation}
where $\Omega_M$ and $\Omega_\Lambda$ are the normalized densities of matter and dark energy today, respectively.

Comparing equations (\ref{eq:Delta_t_DM}) and (\ref{eq:Delta_t_gamma}), we see that $\Delta t_{\rm DM}$ and $\Delta t_{m_\gamma}$ have similar frequency dependence ($\Delta t\propto\nu^{-2}$). Therefore, we can define an effective DM induced by the non-zero photon mass \citep{Shao:2017tuu},
\begin{equation}\label{eq:DM_gamma}
  {\rm DM}_\gamma=\frac{4\pi^2m_e\epsilon_0c^5}{h^2e^2}\frac{H_\gamma(z)}{H_0}m_\gamma^2.
\end{equation}
The observed DM obtained from the dynamical spectrum of FRBs includes both ${\rm DM_{astro}}$ and ${\rm DM}_{\gamma}$, while the former can generally be decomposed into four main parts: the Milky Way interstellar medium (${\rm DM_{MW}}$), the Galactic halo (${\rm DM_{halo}}$), the intergalactic medium (${\rm DM_{IGM}}$), and the host galaxy (${\rm DM_{host}}$). Therefore, the total observed DM of an extragalactic FRB can be written as
\begin{equation}\label{eq:DM_obs}
  {\rm DM_{obs}}={\rm DM_{astro}}+{\rm DM}_{\gamma} = {\rm DM_{MW}}+{\rm DM_{halo}}+{\rm DM_{IGM}}+\frac{{\rm DM_{host}}}{1+z}+{\rm DM}_\gamma,
\end{equation}
where the factor $1+z$ arises from the cosmic expansion.

Thanks to the detailed observation of electron distribution around our Galaxy, the ${\rm DM_{MW}}$ term can be well modeled, such as the NE2001 model \citep{Cordes:2002wz} and the YMW16 model \citep{Yao_2017msh}. These two electron models give consistent results at high Galactic latitude ($b\gtrsim10^{\circ}$), but the YMW16 model may overestimate ${\rm DM_{MW}}$ at low Galactic latitude ($b\lesssim10^{\circ}$) \citep{KochOcker:2021fia}. Therefore, we use the NE2001 model to estimate ${\rm DM_{MW}}$. The ${\rm DM_{halo}}$ term is still poorly known, but it is expected to be in the range $50\sim 80~{\rm pc~cm^{-3}}$ \citep{Prochaska:2019mkd}. Here we take a conservative value ${\rm DM_{halo}}=50~{\rm pc~cm^{-3}}$. Taking a larger value of ${\rm DM_{halo}}$ will lead to a smaller value of ${\rm DM}_\gamma$, thus is expected to give a tighter constraint on photon mass. However, the concrete value of ${\rm DM_{halo}}$ should not significantly affect our results, as it is much smaller than the uncertainties of ${\rm DM_{IGM}}$ and ${\rm DM_{host}}$ terms described below. Therefore, the first two terms on the right-hand-side of equation (\ref{eq:DM_obs}) can be subtracted from the observed ${\rm DM_{obs}}$. For convenience, we define
\begin{equation}\label{eq:DM_E}
  {\rm DM_{obs}'}\equiv {\rm DM_{obs}}-{\rm DM_{MW}}-{\rm DM_{halo}}={\rm DM_{IGM}}+\frac{{\rm DM_{host}}}{1+z}+{\rm DM}_\gamma,
\end{equation}
and treat ${\rm DM_{obs}'}$ as an observable. If the ${\rm DM_{IGM}}$ and ${\rm DM_{host}}$ terms can be modeled properly, then the photon mass can be constrained from FRB data.

In the standard $\Lambda$CDM cosmological model, the mean value of ${\rm DM_{IGM}}$ can be written as \citep{Deng:2013aga,Zhang:2020ass}
\begin{equation}\label{eq:DM_IGM}
  \langle{\rm DM_{IGM}}(z)\rangle=\frac{21cH_0\Omega_bf_{\rm IGM}}{64\pi Gm_p}H_e(z),
\end{equation}
where $f_{\rm IGM}=0.84$ is the fraction of baryon mass in IGM, $m_p$ is the proton mass, $H_0$ is the Hubble constant, $G$ is the Newtonian gravitational constant, $\Omega_b$ is the normalized baryon matter density today. The function $H_e(z)$ is defined by
\begin{equation}\label{eq:H_e}
  H_e(z)=\int_0^z\frac{1+z}{\sqrt{\Omega_m(1+z)^3+\Omega_\Lambda}}dz.
\end{equation}
In this paper, we work in the standard $\Lambda$CDM model with the Planck 2018 parameters, i.e. $H_0=67.4~{\rm km~s^{-1}~Mpc^{-1}}$, $\Omega_m=0.315$, $\Omega_\Lambda=0.685$ and $\Omega_{b}=0.0493$ \citep{Aghanim:2018eyx}.

\begin{figure}
 \centering
 \includegraphics[width=0.6\textwidth]{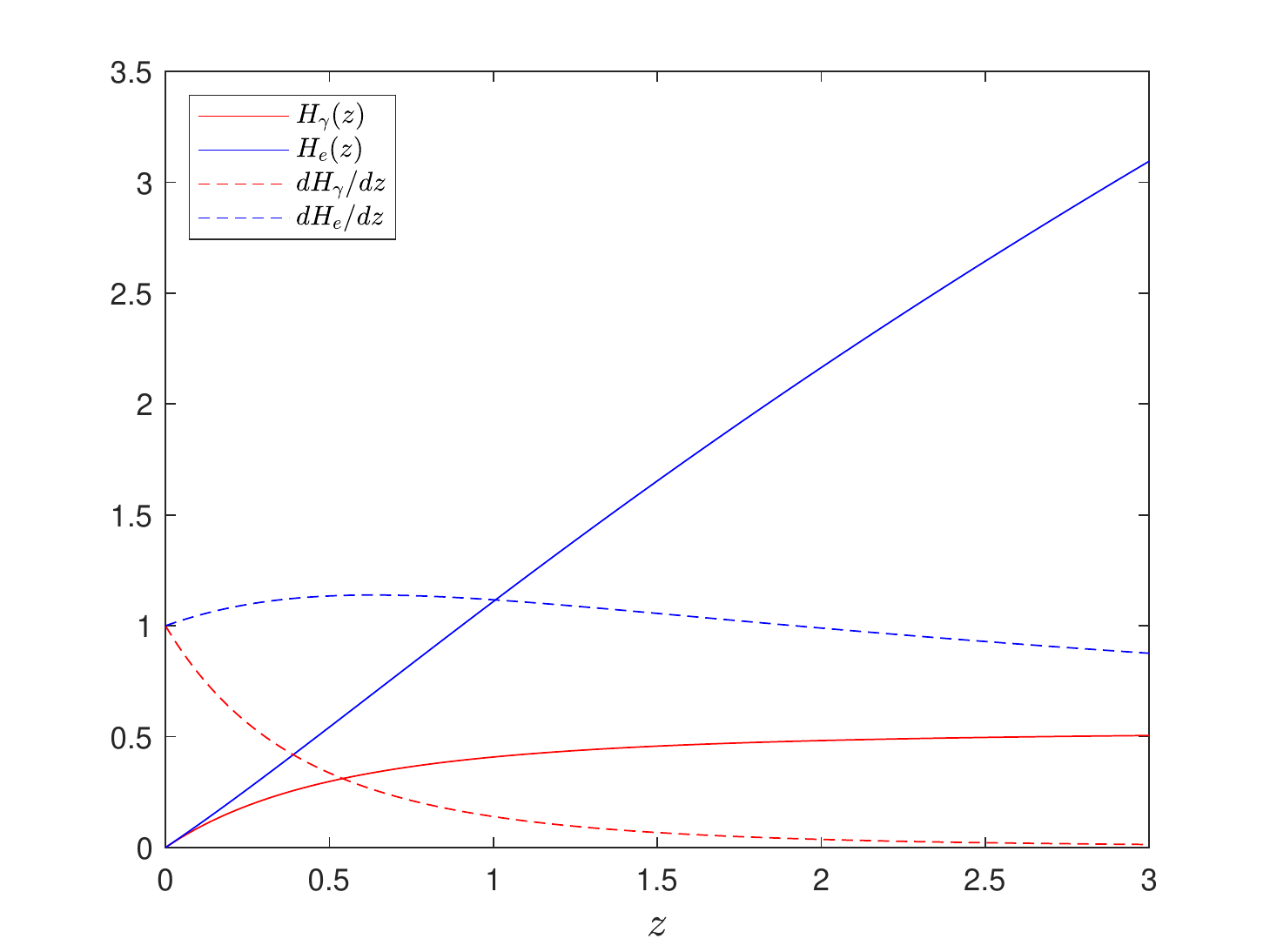}
 \caption{The redshift dependence of $H_e(z)$, $H_\gamma(z)$ and their derivatives.}\label{fig:Hz}
\end{figure}

Comparing equation (\ref{eq:H_gamma}) and equation (\ref{eq:H_e}), we see that the two terms ${\rm DM_{IGM}}$ and ${\rm DM}_\gamma$ have remarkably different redshift dependence. In Figure \ref{fig:Hz}, we plot the redshift dependence of $H_e(z)$ and $H_\gamma(z)$, as well as their derivatives. $H_e(z)$ increases with redshift much faster than $H_\gamma(z)$, since FRBs at higher redshift encounter electrons that are much denser than present day, thus causes the plasma effect to have a strong net increase with redshift. Furthermore, these two terms also differ from the ${\rm DM_{host}}$ term, which evolves with redshift as $(1+z)^{-1}$. Therefore, the degeneracy between ${\rm DM_{IGM}}$, ${\rm DM}_\gamma$ and ${\rm DM_{host}}$ is expected to break down if a large sample of FRBs at different redshift are observed.

It should be emphasized that equation (\ref{eq:DM_IGM}) is the mean value of ${\rm DM_{IGM}}$. The actual value would deviate from the mean caused by e.g. density fluctuation of the Universe. Theoretical analysis and numerical simulations show that the probability of ${\rm DM_{IGM}}$ follows the distribution \citep{Macquart:2020lln,Zhang:2020xoc}
\begin{equation}\label{eq:P_igm}
  p_{\rm IGM}(\Delta)=A\Delta^{-\beta}\exp\left[-\frac{(\Delta^{-\alpha}-C_0)^2}{2\alpha^2\sigma_{\rm IGM}^2}\right], ~~~\Delta>0,
\end{equation}
where $\Delta\equiv{\rm DM_{IGM}}/\langle{\rm DM_{IGM}}\rangle$, $\sigma_{\rm IGM}$ is a deviation parameter, $A$ is a normalization constant, $C_0$ is chosen such that the mean of this distribution is unity, and $\alpha=\beta=3$ \citep{Macquart:2020lln}. Simulations also show that the deviation parameter $\sigma_{\rm IGM}$ is redshift-dependent \citep{McQuinn:2013tmc,Jaroszynski:2018vgh}, so we follow \citet{Macquart:2020lln} and parameterize it as $\sigma_{\rm IGM}=Fz^{-1/2}$, where $F$ is a free parameter.

The host term ${\rm DM_{host}}$ is poorly known, and may vary significantly from bursts to bursts. It is expected to range from several tens to several hundreds ${\rm pc~cm^{-3}}$. For example, \citet{Bannister:2019iju} obtained ${\rm DM_{host}}=30\sim 81~{\rm pc~cm}^{-3}$ for the nonrepeating burst FRB20180924B. \citet{Xu:2021qdn} estimated that the ${\rm DM_{host}}$ of the repeating burst FRB20201124A is in the range $10< {\rm DM_{host}}< 310~{\rm pc~cm}^{-3}$. Theoretical analysis and numerical simulations show that the probability of ${\rm DM_{host}}$ follows the log-normal distribution \citep{Macquart:2020lln,Zhang:2020mgq},
\begin{equation}\label{eq:P_host}
  p_{\rm host}({\rm DM_{host}}|\mu,\sigma_{\rm host})=\frac{1}{\sqrt{2\pi}{\rm DM_{host}}\sigma_{\rm host}} \exp\left[-\frac{(\ln {\rm DM_{host}}-\mu)^2}{2\sigma_{\rm host}^2}\right],
\end{equation}
where $\mu$ and $\sigma_{\rm host}$ are the mean and standard deviation of $\ln {\rm DM_{host}}$, respectively. In general, the two parameters ($\mu,\sigma_{\rm host}$) may be redshift-dependent. Numerical simulations show that the median value of ${\rm DM_{host}}$ (i.e. the exponential of $\mu$) moderately increases with redshift, but $\sigma_{\rm host}$ does not vary significantly with redshift \citep{Zhang:2020mgq}. Therefore, we treat $\sigma_{\rm host}$ as a constant, and parameterize $\mu$ with the power-law form \citep{Zhang:2020mgq},
\begin{equation}
  e^\mu=e^{\mu_0}(1+z)^\varepsilon.
\end{equation}
Thus the probability distribution $p_{\rm host}$ have three parameters ($\mu_0,\sigma_{\rm host},\varepsilon$) in total.

Given the probability distributions $p_{\rm IGM}$ and $p_{\rm host}$, we can calculate the probability distribution of ${\rm DM_{obs}'}$ at any redshift $z$, which can be written according to equation (\ref{eq:DM_E}) as
\begin{equation}\label{eq:P_E}
  p_{\rm obs}({\rm DM_{\rm obs}'}|z)=\int_0^{(1+z)(\rm DM_{\rm obs}'-{\rm DM_\gamma})} p_{\rm host}({\rm DM_{host}}|\mu_0,\sigma_{\rm host},\varepsilon) p_{\rm IGM}({\rm DM_{\rm obs}'}-\frac{\rm DM_{host}}{1+z}-{\rm DM_\gamma}|F,z)d{\rm DM_{host}}.
\end{equation}
If a sample of FRBs are observed, the likelihood function is given by
\begin{equation}
  \mathcal{L}({\rm FRBs}|F,\mu_0,\sigma_{\rm host},\varepsilon,m_\gamma)=\prod_{i=1}^Np_{\rm obs}({\rm DM_{\rm obs,\it i}'}|z_i).
\end{equation}
According to Bayes theorem, the posterior probability distribution of the free parameters ($F,\mu_0,\sigma_{\rm host},\varepsilon,m_\gamma$) is given by
\begin{equation}
  P(F,\mu_0,\sigma_{\rm host},\varepsilon,m_\gamma|{\rm FRBs})\propto\mathcal{L}({\rm FRBs}|F,\mu_0,\sigma_{\rm host},\varepsilon,m_\gamma)P_0(F,\mu_0,\sigma_{\rm host},\varepsilon,m_\gamma),
\end{equation}
where $P_0$ is the prior of the parameters.

\section{Data and Results}\label{sec:results}

So far, there are in total 19 extragalactic FRBs that have direct measurement of redshift\footnote{The FRB Host Database, http://frbhosts.org/}. Among them, we ignore FRB20200120E and FRB20190614D. The former is so close to our Galaxy ($D\approx 3.6$ Mpc) that its redshfit is negtive ($z=-0.0001$)\footnote{The redshift is dominated by peculiar velocity, rather than the Hubble flow.} \citep{Bhardwaj:2021xaa,Kirsten:2021llv}, while the latter has no measurement of spectroscopic redshift, but has photometric redshift ($z_{\rm ph}\approx 0.6$) \citep{Law:2020cnm}. All the rest 17 FRBs have well measured spectroscopic redshift. The main properties of the 17 FRBs are listed in Table \ref{tab:host}, which will be used to constrain photon mass.

\begin{table}
\centering
\caption{\small{The main properties of 17 well-localized FRBs. ${\rm DM_{MW}}$ is calculated using the NE2001 model, and ${\rm DM_{obs}'}$ is calculated by subtracting ${\rm DM_{\rm MW}}$ and ${\rm DM_{\rm halo}}$ from the observed ${\rm DM_{\rm obs}}$, assuming ${\rm DM_{\rm halo}}=50~{\rm pc~cm^{-3}}$.}}\label{tab:host}
{\begin{tabular}{ccccccccl} 
\hline\hline 
FRBs & RA & Dec & ${\rm DM_{obs}}$ & ${\rm DM_{MW}}$ & ${\rm DM_{obs}'}$ & $z_{\rm sp}$ & repeat? & reference\\
& [ $^{\circ}$ ] & [ $^{\circ}$ ] & [${\rm pc~cm^{-3}}$] & [${\rm pc~cm^{-3}}$] & [${\rm pc~cm^{-3}}$] & & \\
\hline
20121102A & $82.99$ & $33.15$ &557.00 &157.60 &349.40 &0.1927 & Yes & \citet{Chatterjee:2017dqg}\\
20180301A & $93.23$ & $4.67$ &536.00 &136.53 &349.47 &0.3305 & Yes & \citet{Bhandari:2021pvj}\\
20180916B & $29.50$ & $65.72$ &348.80 &168.73 &130.07 &0.0337 & Yes & \citet{Marcote:2020ljw}\\
20180924B & $326.11$ & $-40.90$ &362.16 &41.45 &270.71 &0.3214 & No & \citet{Bannister:2019iju}\\
20181030A & $158.60$ & $73.76$ &103.50 &40.16 &13.34 &0.0039 & Yes & \citet{Bhardwaj:2021hgc}\\
20181112A & $327.35$ & $-52.97$ &589.00 &41.98 &497.02 &0.4755 & No & \citet{2019Sci...366..231P}\\
20190102C & $322.42$ & $-79.48$ &364.55 &56.22 &258.33 &0.2913 & No & \citet{Macquart:2020lln}\\
20190523A & $207.06$ & $72.47$ &760.80 &36.74 &674.06 &0.6600 & No & \citet{Ravi:2019alc}\\
20190608B & $334.02$ & $-7.90$ &340.05 &37.81 &252.24 &0.1178 & No & \citet{Macquart:2020lln}\\
20190611B & $320.74$ & $-79.40$ &332.63 &56.60 &226.03 &0.3778 & No & \citet{Macquart:2020lln}\\
20190711A & $329.42$ & $-80.36$ &592.60 &55.37 &487.23 &0.5217 & Yes & \citet{Macquart:2020lln}\\
20190714A & $183.98$ & $-13.02$ &504.13 &38.00 &416.13 &0.2365 & No & \citet{2020ApJ...903..152H}\\
20191001A & $323.35$ & $-54.75$ &507.90 &44.22 &413.68 &0.2340 & No & \citet{2020ApJ...903..152H}\\
20191228A & $344.43$ & $-29.59$ &297.50 &33.75 &213.75 &0.2432 & No &\citet{Bhandari:2021pvj}\\
20200430A & $229.71$ & $12.38$ &380.25 &27.35 &302.90 &0.1608 & No & \citet{Bhandari:2021pvj}\\
20200906A & $53.50$ & $-14.08$ &577.80 &36.19 &491.61 &0.3688 & No & \citet{Bhandari:2021pvj}\\
20201124A & $77.01$ & $26.06$ &413.52 &126.49 &237.03 &0.0979 & Yes &\citet{Fong:2021xxj}\\
\hline
\end{tabular}}
\end{table}

\begin{figure}
 \centering
 \includegraphics[width=0.48\textwidth]{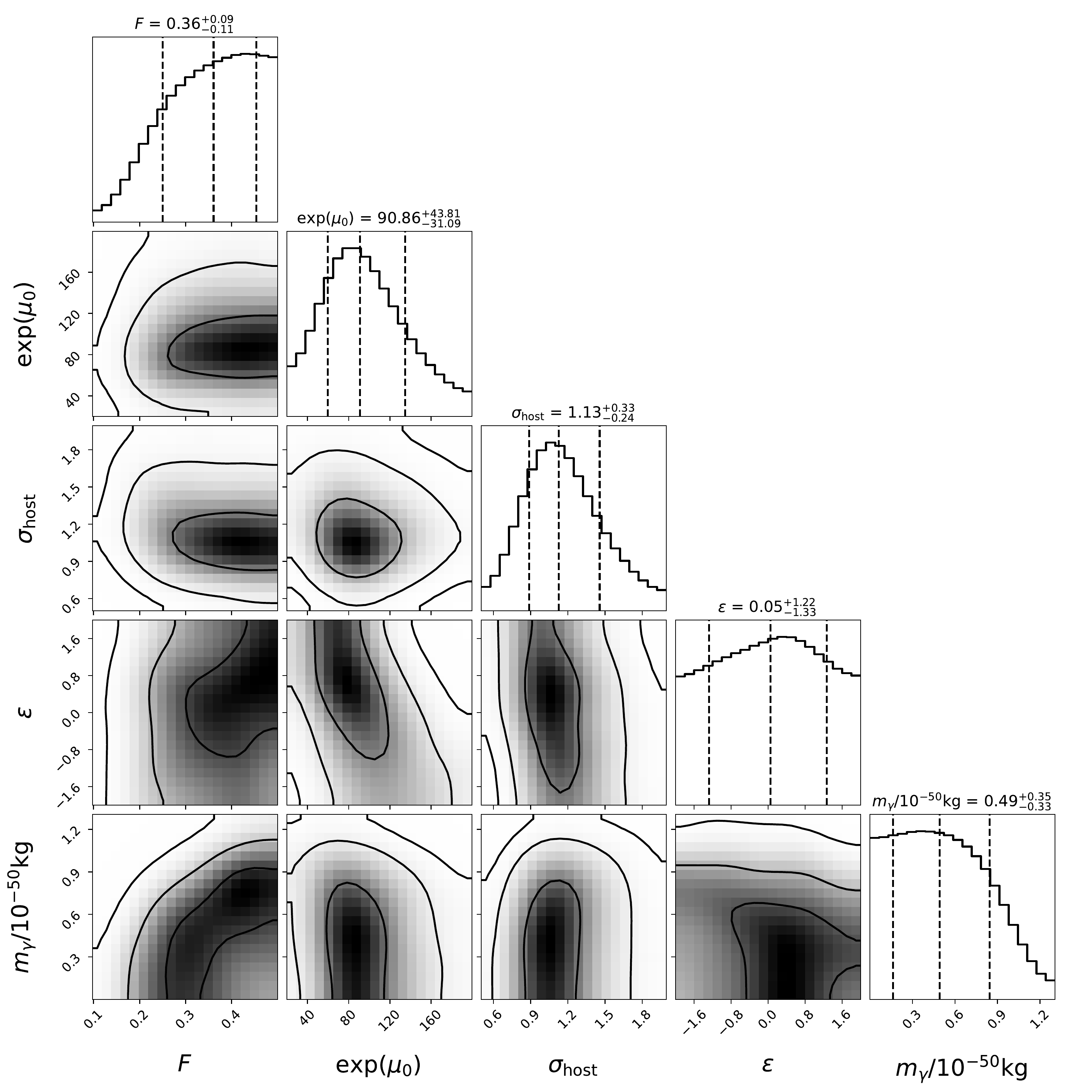}
 \includegraphics[width=0.48\textwidth]{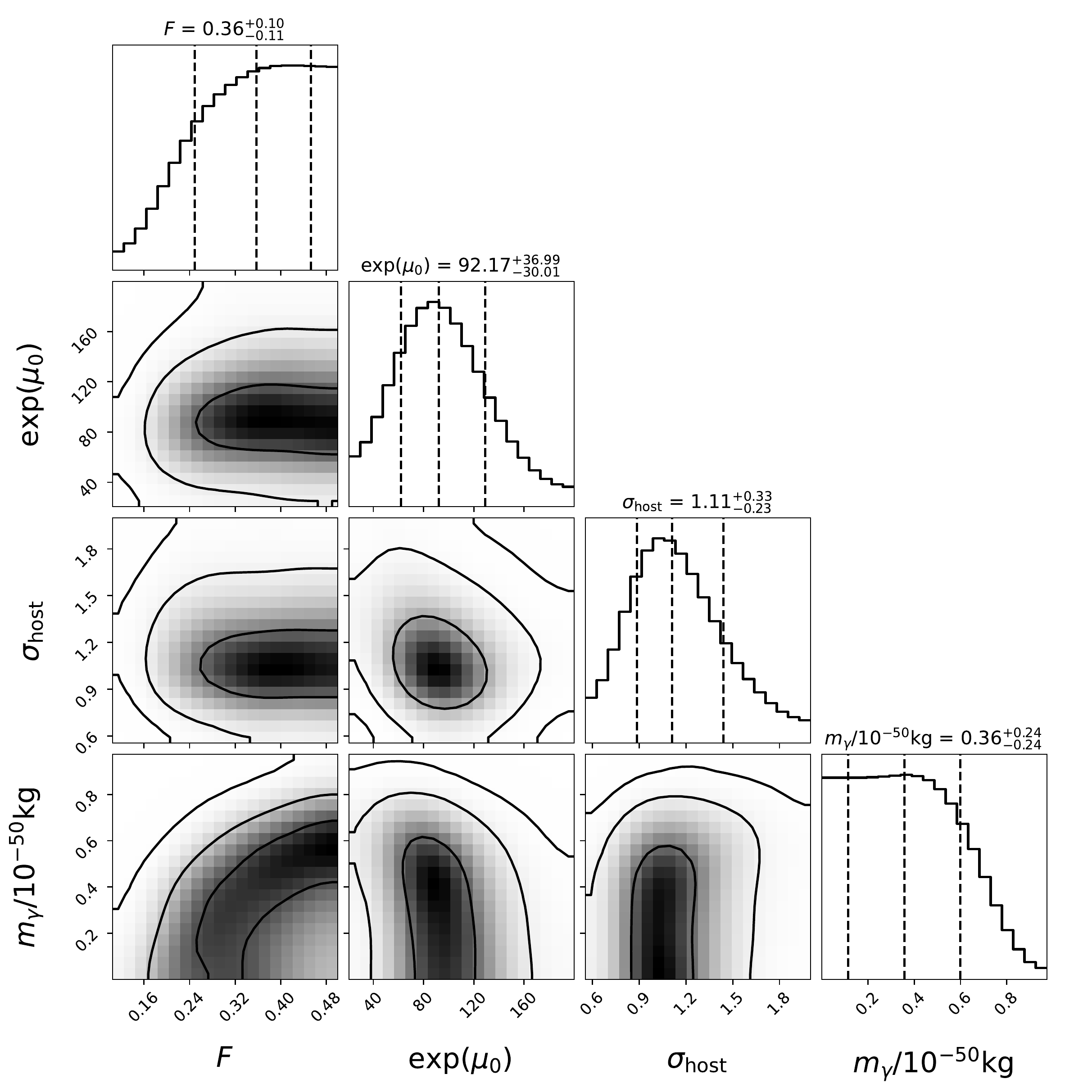}
 \caption{The contour plot of the free parameters ($F, e^{\mu_0}, \sigma_{\rm host}, \varepsilon, m_\gamma$) constrained from 17 well-localized FRBs. The contours from the inner to outer represent $1\sigma$, $2\sigma$ and $3\sigma$ confidence regions, respectively. Left panel: $\varepsilon$ is a free parameter; Right panel: $\varepsilon=0$ fixed.}\label{fig:contour}
\end{figure}

We constrain the free parameters ($F,\mu_0,\sigma_{\rm host},\varepsilon,m_\gamma$) simultaneously use 17 well-localized FRBs. In practice, we use $e^{\mu_0}$ instead of ${\mu_0}$ as a free parameter, because the former directly represents the median value of ${\rm DM_{host}}$ (at $z=0$). The posterior probability density functions of the free parameters are calculated with the Markov Chain Monte Carlo method using the publicly available python code \textsf{emcee} \citep{Foreman-Mackey:2012any}, while the other cosmological parameters are fixed to the Planck 2018 values \citep{Aghanim:2018eyx}. Flat priors are adopted for all the free parameters: $F\in\mathcal{U}(0.01,0.5)$, $e^{\mu_0}\in \mathcal{U}(20,200)~{\rm pc~cm^{-3}}$, $\sigma_{\rm host}\in \mathcal{U}(0.2,2.0)$, $\varepsilon\in\mathcal{U}(-2,2)$, and $m_\gamma\in \mathcal{U}(10^{-69},10^{-42})$ kg\footnote{The lower limit of $m_\gamma$ is determined by the uncertainty principle $m_\gamma c^2T\gtrsim h$, where $T\approx 10^{10}$ years is the age of the Universe. The upper limit of $m_\gamma$ is determined by the requirement that $m_\gamma c^2\lesssim h\nu$, where $\nu\approx 1$ GHz is the frequency of FRB emission.}. We report the median values and the $1\sigma$ uncertainties of the parameters in Table $\ref{tab:results}$. For the photon mass, the $1\sigma$ and $2\sigma$  upper limits are reported. The posterior probability density functions and the confidence contours of the parameters are plotted in the left panel of Figure \ref{fig:contour}. The parameters $F$, $e^{\mu_0}$ and $\sigma_{\rm host}$ can be tightly constrained, with $F=0.36_{-0.11}^{+0.09}$, $e^{\mu_0}=90.86_{-31.09}^{+43.81}~ {\rm pc~cm^{-3}}$ and $\sigma_{\rm host}=1.13_{-0.24}^{+0.33}$. However, the parameter $\varepsilon$ can't be tightly constrained, with $\varepsilon=0.05_{-1.33}^{+1.22}$. The $1\sigma$ ($2\sigma$) upper limit of photon mass is constrained to be $m_\gamma<0.67\times 10^{-50}$ kg ($m_\gamma<1.01\times 10^{-50}$ kg).

\begin{table}
\centering
\caption{\small{The parameters $(F,e^{\mu_0},\sigma_{\rm host},\varepsilon,m_\gamma)$ constrained from 17 well-localized FRBs. The uncertainties are given at $1\sigma$ confidence level. For the photon mass, the $1\sigma$ and $2\sigma$  upper limits are reported.}}\label{tab:results}
{\begin{tabular}{ccccc} 
\hline\hline 
F &$e^{\mu_0}/{\rm pc~cm^{-3}}$  &  $\sigma_{\rm host}$  & $\varepsilon$ & $m_\gamma/10^{-50}~{\rm kg}$ \\
\hline
$0.36_{-0.11}^{+0.09}$ & $90.86_{-31.09}^{+43.81}$ & $1.13_{-0.24}^{+0.33}$ & $0.05_{-1.33}^{+1.22}$  & $<0.67~(<1.01)$\\
$0.36_{-0.11}^{+0.10}$ & $92.17_{-30.01}^{+36.99}$ & $1.11_{-0.23}^{+0.33}$ & $\varepsilon=0$ (fixed) & $<0.48~(<0.71)$\\
\hline
\end{tabular}}
\end{table}

As we can see, $\varepsilon$ is well consistent with zero, implying no strong evidence for the redshift evolution of ${\rm DM_{host}}$. Therefore, we fix $\varepsilon=0$ and fit the remaining four parameters ($F,e^{\mu_0},\sigma_{\rm host},m_\gamma$). The best-fitting results are: $F=0.36_{-0.11}^{+0.10}$, $e^{\mu_0}=92.17_{-30.01}^{+36.99}~{\rm pc~cm^{-3}}$, $\sigma_{\rm host}=1.11_{-0.23}^{+0.33}$, and the $1\sigma$ ($2\sigma$) upper limit of photon mass is $m_\gamma<0.48\times 10^{-50}$ kg ($m_\gamma<0.71\times 10^{-50}$ kg). The posterior probability density functions and the confidence contours of the parameters are plotted in the right panel of Figure \ref{fig:contour}. As is seen, the 4-parameter fit and the 5-parameter fit give consistent results, further confirming that there is no obvious redshift evolution of ${\rm DM_{host}}$.

\section{Monte Carlo Simulations}\label{sec:simulations}

With the running of new radio telescopes, such as the Canadian Hydrogen Intensity Mapping Experiment (CHIME, \citealt{Amiri:2018qsq}) and the Five-hundred-meter Aperture Spherical Telescope (FAST, \citealt{Nan:2011um}), it is expected that more and more FRBs can be discovered in the next years, among which several of them can be well localized. Therefore, it is interesting to investigate the constraining ability on photon mass if more well-localized FRBs are available in the future. To this end, we perform Monte Carlo simulations.

Due do the lack of direct redshift measurement, the intrinsic redshift distribution of FRBs is still unclear. Several possibilities have been discussed in literatures. For example, \citet{Yu:2017beg}  assumed that the redshift distribution of FRBs is similar to that of GRBs, \citet{Li:2019klc} assumed that FRBs have a constant comoving number density but with a Gaussian cutoff, and \citet{Zhang:2020ass} discussed that the intrinsic event rate density of FRBs tracks the SFR, or relates to the compact star merger but with an additional time delay. Here we adopt the SFR-related redshift distribution, which takes the form \citep{Zhang:2020ass}
\begin{equation}\label{eq:pdf_redshift}
  P(z)\propto\frac{4\pi D^2_c(z){\rm SFR}(z)}{(1+z)H(z)},
\end{equation}
where $D_c(z)=\int_0^zc/H(z)dz$ is the comoving distance, $H(z)=H_0\sqrt{\Omega_m(1+z)^3+\Omega_\Lambda}$ is the Hubble expansion rate, and the SFR evolves with redshift as \citep{Yuksel:2008cu}
\begin{equation}\label{eq:SFR}
  {\rm SFR}(z)=0.02\left[(1+z)^{a\eta}+\left(\frac{1+z}{B}\right)^{b\eta}+\left(\frac{1+z}{C}\right)^{c\eta}\right]^{1/\eta},
\end{equation}
where $a=3.4$, $b=-0.3$, $c=-3.5$, $B=5000$, $C=9$ and $\eta=-10$.

We simulate a mock sample of FRBs, each FRB contains the parameters ($z,{\rm DM_{obs}'}$). The simulations are performed based on the standard $\Lambda$CDM model with the Planck2018 parameters: $H_0=67.4~{\rm km~s^{-1}~Mpc^{-1}}$, $\Omega_m=0.315$, $\Omega_\Lambda=0.685$ and $\Omega_b=0.0493$ \citep{Aghanim:2018eyx}. The other fiducial parameters are $F=0.3$, $e^{\mu_0}=100~{\rm pc~cm^{-3}}$, $\sigma_{\rm host}=1$,  $\varepsilon=0$, and $m_\gamma=0$. As we are interested in the constraint on photon mass, rather than the redshift evolution of ${\rm DM_{host}}$, we just treat $e^\mu$ as a constant (i.e. fix $\varepsilon=0$). This is reasonable at present, since there is no evidence for the redshift evolution of $e^\mu$ in the real FRB data. The procedures of simulation are as follows:
\begin{enumerate}[(1)]
  \item{Randomly draw redshift $z$ from the probability distribution in equation (\ref{eq:pdf_redshift}), with the upper limit $z_{\rm max}=3$.}
  \item{Calculate the mean value $\langle{\rm DM_{IGM}}(z)\rangle$ according to equation (\ref{eq:DM_IGM}).}
  \item{Randomly draw ${\rm DM_{IGM}}$ from the probability distribution in equation (\ref{eq:P_igm}).}
  \item{Randomly draw ${\rm DM_{host}}$ from the probability distribution in equation (\ref{eq:P_host}).}
  \item{Calculate ${\rm DM_{obs}'}$ according to equation (\ref{eq:DM_E}), where ${\rm DM}_\gamma=0$ (zero photon mass).}
\end{enumerate}

\begin{figure}
 \centering
 \includegraphics[width=0.48\textwidth]{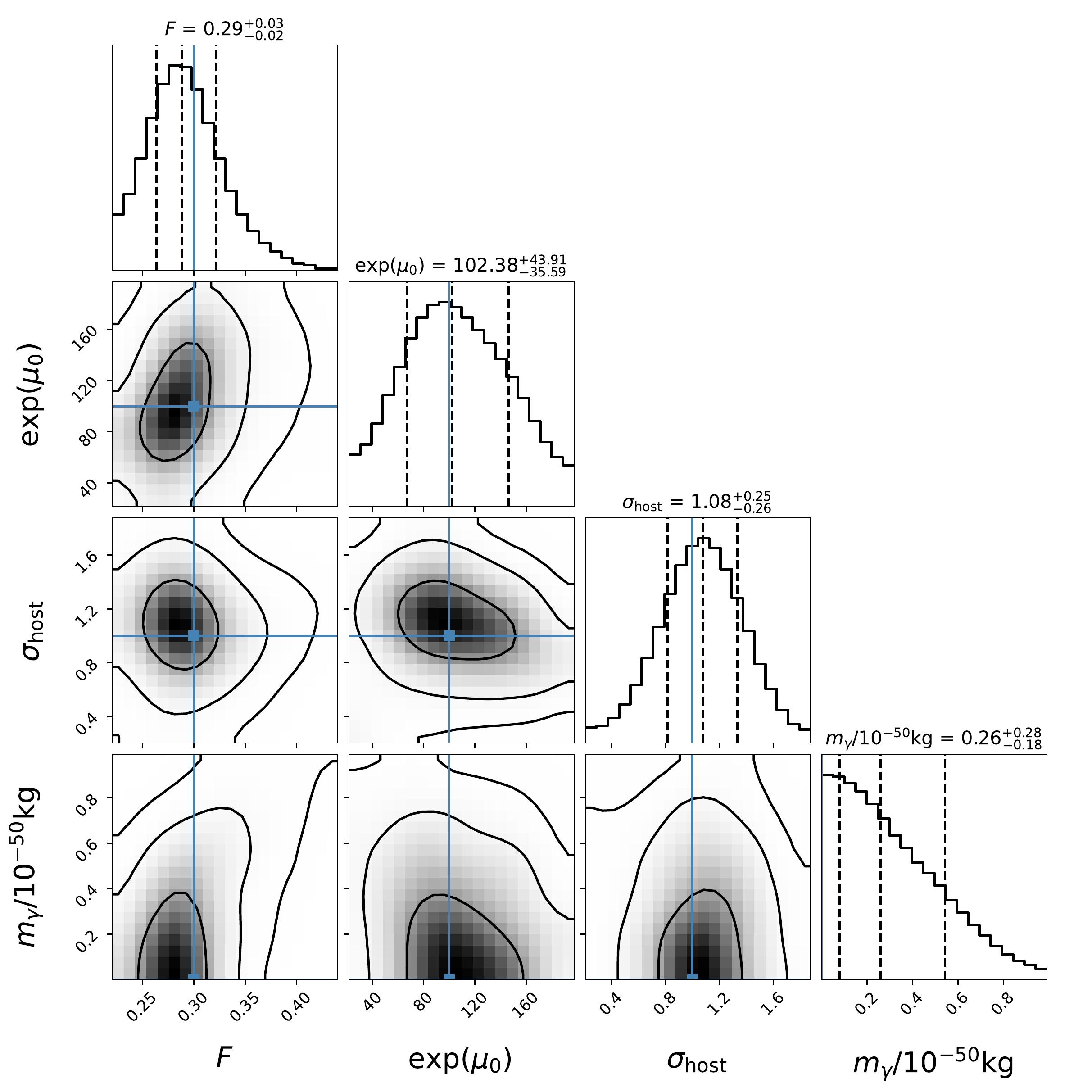}
 \includegraphics[width=0.48\textwidth]{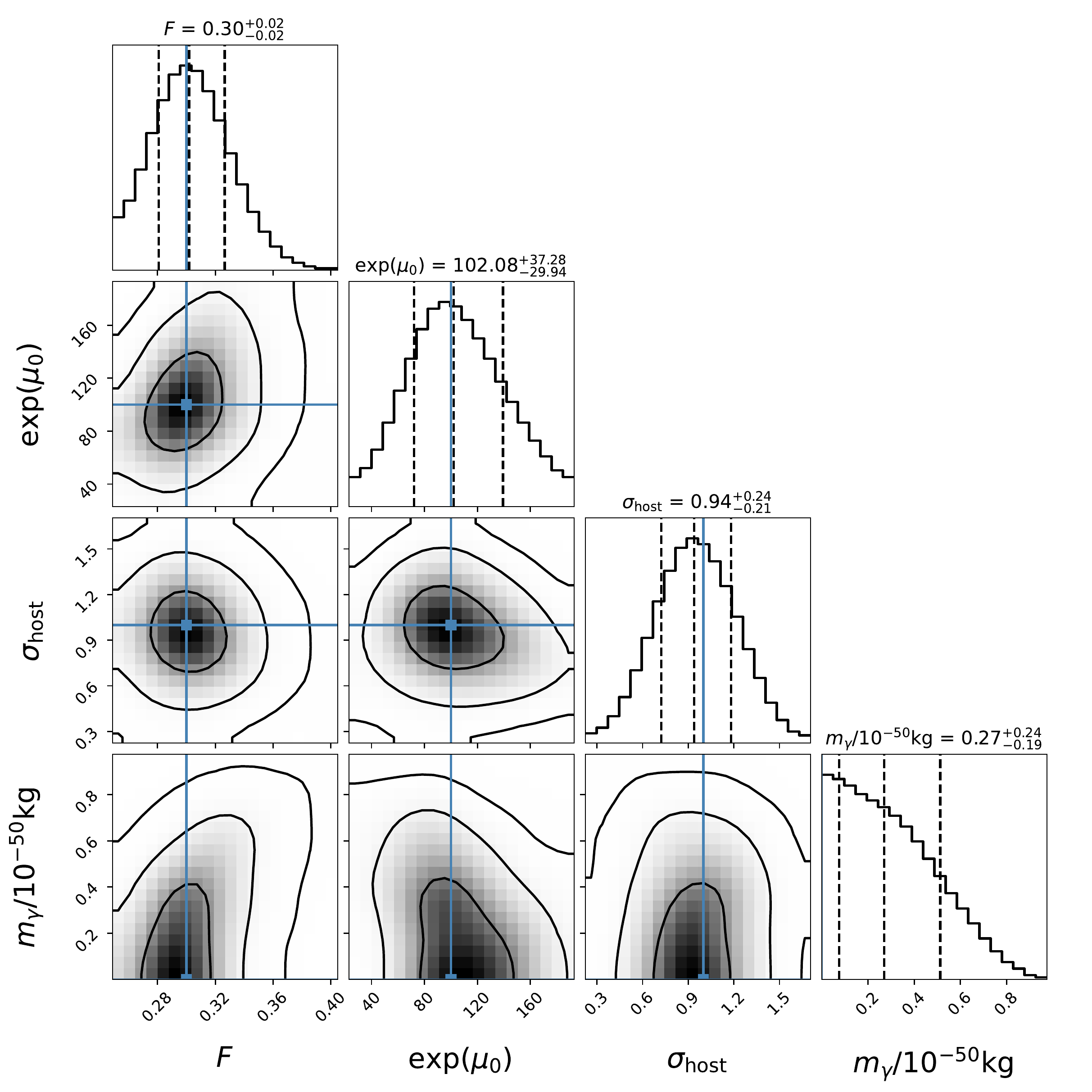}
 \caption{The contour plot of the free parameters ($F, e^{\mu_0}, \sigma_{\rm host}, m_\gamma$) constrained from different number of mock FRBs (left panel: N=100; right panel: N=200). The vertical dashed lines from left to right in each subfigures represent the 16\%, 50\% and 84\% quantiles of the distributions, respectively. The solid lines represent the fiducial values. The contours from the inner to outer represent $1\sigma$, $2\sigma$ and $3\sigma$ confidence regions, respectively.}\label{fig:simulation}
\end{figure}

With the mock FRB sample, we constrain photon mass (together with other parameters $F$, $e^{\mu_0}$ and $\sigma_{\rm host}$) using the method described in Section \ref{sec:methodology}. In Figure \ref{fig:simulation}, we shows the contour plots of parameter space in two arbitrary realizations of simulation with different number of FRBs (left panel: $N=100$; right panel: $N=200$). We see that the best-fitting parameters $F$, $e^{\mu_0}$ and $\sigma_{\rm host}$ can correctly recover the fiducial values within $1\sigma$ uncertainty, and the upper limit of photon mass can be tightly constrained. Compared with the real data, the precision of the parameter $F$ is highly improved. However, the precision of the other parameters does not change significantly. We also note that enlarging the FRB sample from 100 to 200 does not significantly improve the constraining ability on photon mass.

\begin{figure}
 \centering
 \includegraphics[width=\textwidth]{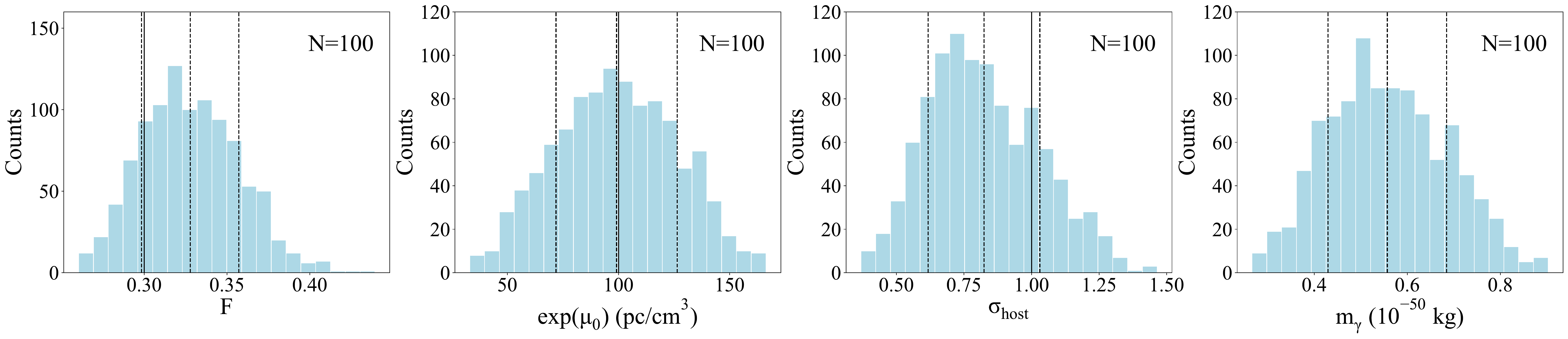}
 \includegraphics[width=\textwidth]{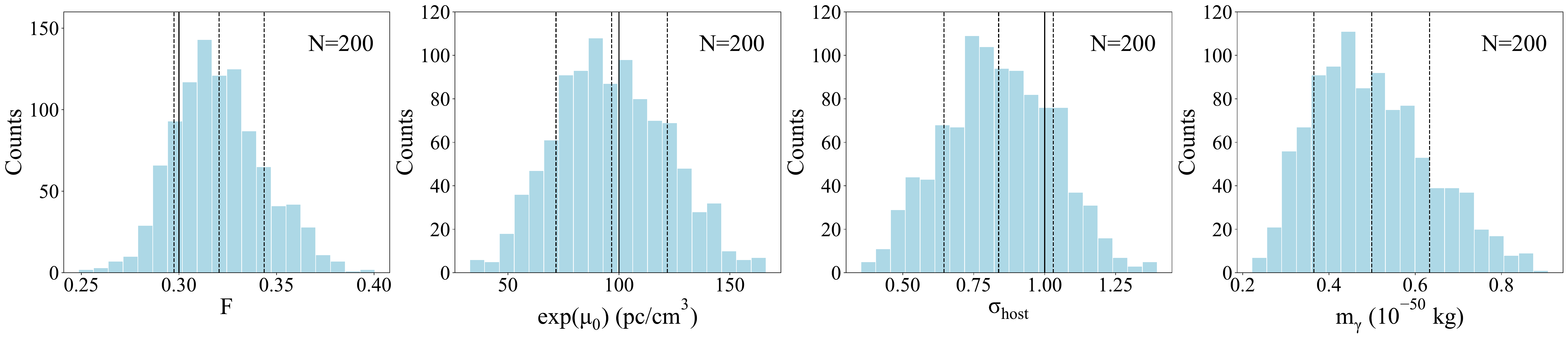}
 \caption{The distributions of the best-fitting parameters in 1000 simulations. The vertical dashed lines from left to right in each subfigures represent the 16\%, 50\% and 84\% quantiles of the distributions, respectively. The vertical solid lines represent the fiducial values. Top panels: N=100; bottom panels: N=200.}\label{fig:simulation1000}
\end{figure}

Considering the statistical fluctuation, we simulate 1000 times for a fixed number of FRBs. In each simulation, we calculate the median value of the parameters ($F, e^{\mu_0},\sigma_{\rm host}$), and the $1\sigma$ upper limit of the photon mass ($m_\gamma$). The distributions of the parameters in 1000 simulations are shown in Figure \ref{fig:simulation1000} (top panels for $N=100$ and bottom panels for $N=200$). We see that the distribution of each parameter is approximately Gaussian. We calculate the 16\%, 50\% and 84\% quantiles of each distribution and list the results in Table \ref{tab:simulation} (also shown in Figure \ref{fig:simulation1000} as the dashed lines). We can see that the fiducial value of each parameter ($F, e^{\mu_0},\sigma_{\rm host}$) falls into the $1\sigma$ confidence region of the distribution. We note that the distribution of $e^{\mu_0}$ is well consistent with the fiducial value, but the parameters $F$ and $\sigma_{\rm host}$ are a little biased. The median value of $F$ is larger than the fiducial value. On the contrary, the median value of $\sigma_{\rm host}$ is smaller than the fiducial value. This may be caused by the correlation between parameters. Nevertheless, all the parameters are consistent with the fiducial values within $1\sigma$ uncertainty. With 100 and 200 FRBs, the photon mass can be constrained at the level of $m_\gamma<0.55\times 10^{-50}$ kg and $m_\gamma<0.49\times 10^{-50}$ kg, respectively. This further confirms that increasing the number of FRBs can't significantly tighten the constraint on photon mass, and the possible reasons will be discussed in the next section.

\begin{table}
\centering
\caption{\small{The median values and $1\sigma$ confidence regions of the parameters ($F, e^{\mu_0},\sigma_{\rm host},m_\gamma$) in 1000 simulations. The fiducial parameters used in the simulations are $F=0.3$, $e^{\mu_0}=100~{\rm pc~cm~^{-3}}$, $\sigma_{\rm host}=1$ and $m_\gamma=0$.}}\label{tab:simulation}
{\begin{tabular}{ccccc} 
\hline\hline 
$N$ & $F$ &  $e^{\mu_0}/{\rm pc~cm^{-3}}$  &  $\sigma_{\rm host}$  & $m_\gamma/10^{-50}~{\rm kg}$ \\
\hline
100 & $0.32_{-0.03}^{+0.03}$ & $98.90_{-28.62}^{+28.10}$ & $0.80_{-0.19}^{+0.24}$ & $0.55_{-0.12}^{+0.15}$\\
200 & $0.32_{-0.03}^{+0.03}$ & $94.88_{-23.28}^{+28.16}$ & $0.83_{-0.19}^{+0.20}$ & $0.49_{-0.12}^{+0.16}$\\
\hline
\end{tabular}}
\end{table}

\section{Discussion and Conclusions}\label{sec:conclusions}

In this paper, we constrained the photon mass from a sample of well-localized FRBs using Bayes inference method. The probability distributions of DMs of IGM and host galaxy were properly taken into account. The degeneracy between time delays induced by DM and non-zero photon mass was broken thanks to the different redshift dependence. The $1\sigma$ and $2\sigma$ upper limits of photon mass were tightly constrained to be $m_\gamma < 4.8 \times 10^{-51}$ kg and $m_\gamma < 7.1\times 10^{-51}$ kg, respectively. Moreover, we also investigated the possible redshift evolution of host DM, but no strong evidence was found. Monte Carlo simulations showed that, the constraint on photon mass can be significantly tightened even if we enlarge the FRB sample to 200 and extend the redshift range up to $z=3$. One possible reason is that the uncertainty on the ${\rm DM_{IGM}}$ is large at high redshift\footnote{Although the relative uncertain $\sigma_{\rm IGM}\propto z^{-1/2}$ decreases with the increasing of redshift, the absolute uncertainty ($\sigma_{\rm IGM}{\rm DM_{IGM}}$) increases with redshift.}. This means that the distribution of ${\rm DM_{IGM}}$ spreads to a wide range at high redshift. Therefore, when marginalizing over the probability distribution of other parameters (including ${\rm DM_{IGM}}$), the posterior probability distribution of photon mass can't be tightened significantly. This can also be see from Figure \ref{fig:contour} and Figure \ref{fig:simulation}, as enlarging the FRB sample from 17 (the real data) to 100 or 200 (the mock data) mainly tightens the constraint on $F$, but the precision of other parameters is not changed significantly. Another reason is that the ${\rm DM}_\gamma$ term is insensitive to redshift. As is seen from Figure \ref{fig:Hz}, $H_\gamma(z)$ is approximately redshift-independent at $z>1$, while $H_e(z)$ evolves with redshift fast at high redshift. This is the reason why extending the redshift range of FRB sample mainly tighten the constraints on $F$, but not the photon mass.

In fact, FRBs have been already widely used to constrain photon mass. Using a single FRB 150418 at redshift $z=0.492$, \citet{Wu:2016brq} obtained the $1\sigma$ upper limits on photon mass $m_\gamma < 5.2\times 10^{-50}$. With the combination of FRB 121102 at redshift $z=0.19273$ and 20 FRBs without direct redshift measurement, \citet{Shao:2017tuu} obtained $m_\gamma < 8.7\times 10^{-51}$ kg. With a sample of nine well-localized FRBs in the redshift range $0<z<0.66$, \citet{Wei:2020wtf} obtained $m_\gamma < 7.1\times 10^{-51}$ kg. In this paper, we constrained photon mass using 17 well-localized FRBs in the redshift range $0<z<0.66$, and obtained $m_\gamma<4.8\times 10^{-51}$ kg. Our constraint on photon mass is a little tighter than that of \citet{Wei:2020wtf}, which is due to the enlargement of FRB sample (17 FRBs versus 9 FRBs). However, the photon mass is constrained at the same order of magnitude ($\sim 10^{-51}$ kg), which is because the FRB sample used in our paper and that used in \citet{Wei:2020wtf} fall into the same redshift range ($0<z<0.66$).

With the constrained photon mass $m_\gamma<4.8\times 10^{-51}$ kg, the effective DM induced by the non-zero photon mass can be estimated according to equation \ref{eq:DM_gamma}. At the maximum redshift $z=0.66$, we have ${\rm DM}_\gamma\lesssim 8$ pc cm$^{-3}$, and $\langle{\rm DM_{IGM}}\rangle\approx 600$ pc cm$^{-3}$. We see that ${\rm DM}_\gamma$ is only about one percent of $\langle{\rm DM_{IGM}}\rangle$. At redshift $z=3$, we have ${\rm DM}_\gamma\lesssim 12$ pc cm$^{-3}$, which is much smaller than $\langle{\rm DM_{IGM}}\rangle\approx 2500$ pc cm$^{-3}$ at the same redshift. Therefore, the DM induced by the non-zero photon mass is in general negligible.

\section*{Acknowledgements}
This work has been supported by the National Natural Science Fund of China (Grant Nos. 11873001, 12147102 and 12275034), and the Fundamental Research Funds for the Central Universities of China (Grants No. 2022CDJXY-002).

\section*{Data Availability}
The Host/FRB catalog is available at the FRB Host Database \textcolor{blue}{http://frbhosts.org}.

\bibliographystyle{mnras}
\bibliography{reference}

\label{lastpage}

\end{document}